\documentclass[prd,twocolumn,showpacs,amsmath,amssymb]{revtex4}
\usepackage{graphicx}
\usepackage{dcolumn}
\usepackage{bm}
\begin{document}
\title{Dynamics of tachyon field with an inverse square potential in loop quantum cosmology}
\author{Fei Huang}
\affiliation{Department of Physics, Beijing Normal University, Beijing 100875, China}
\author{Kui Xiao}
 \email{87xiaokui@mail.bnu.edu.cn}
  \affiliation{Department of Basic Teaching, Hunan Institute of Technology, Hengyang 421002, China}
\author{Jian-Yang Zhu}
\thanks{Author to whom correspondence should be addressed}
 \email{zhujy@bnu.edu.cn}
  \affiliation{Department of Physics, Beijing Normal University, Beijing 100875, China}
\date{\today}
\begin{abstract}
The dynamical behavior of tachyon field with an inverse potential is investigated in loop quantum cosmology. It reveals that the late time behavior of tachyon field with this potential leads to a power-law expansion. In addition, an additional barotropic perfect fluid with the adiabatic index $0<\gamma<2$ is added, and the dynamical system is shown to be an autonomous one. The stability of this autonomous system is discussed using phase plane analysis. There exist up to five fixed points with only two of them possibly stable. The two stable node (attractor) solutions are specified, and their
cosmological indications are discussed. For the tachyon dominated solution, the further discussion is stretched to the possibility of considering tachyon field as a combination of two parts which respectively behave like dark matter and dark energy.
\end{abstract}

\maketitle

\section{Introduction}

The tachyon field with various potential has been studied a lot in
cosmology. Many models have been built by treating tachyon field as inflaton
field \cite{xiong}, candidate of dark energy \cite{Cope05,Cope06}, or a dual
role of the two \cite{sami}. But Ref.\cite{sami} also argued that, for the class
of potentials which $V(\phi)\rightarrow0$ as $\phi\rightarrow\infty$,
radiation domination will never commence since the tachyon field energy
density $\rho_{\phi}$ can at best scale as $a^{-3}$. The radiation energy
density would always redshift faster than the tachyon field. The tachyon
field with an inverse square potential is shown to be able to produce a
power-law expansion \cite{Feinstein}. Coupled with a barotropic perfect
fluid, the dynamical behavior of this potential has been studied in
classical cosmology\cite{juan}. However, the tracking solution, in which the
energy density of the field and the barotropic fluid scales as a same power
of $a$, may not be viable because its constraints on the adiabatic index $%
\gamma$.

Our work of tachyon field cosmology is performed under the framework of loop
quantum cosmology. LQC is a canonical quantization of homogeneous spacetime
using the techniques developed in loop quantum gravity (LQG). The loop
quantum effects can be very well described by the effective theory of LQC. A
modified Friedmann equation is proposed and two corrections are often
considered: the inverse volume correction and the holonomy correction.
However, the holonomy correction dominates over the inverse volume
correction for a universe with a large scale factor, and thus the latter can
be neglected without harm. Therefore, we only consider the holonomy
correction in this paper.

Currently, tachyon matter has not been thoroughly investigated in loop
quantum cosmology(LQC). A. A. Sen \cite{AASEN} generalized the description
of tachyon matter in standard cosmology to LQC under inverse volume
correction. Xiong and Zhu \cite{xiong} investigated the
inflation scenario of a pure tachyon field with an exponential potential
under the holonomy correction. Xiao and Zhu \cite{kui2}
performed a phenomenological analysis of tachyon warm inflation, in which
the tachyon field with an exponential potential is coupled with radiation
and interaction between the two matters were considered. However, the
dynamics of tachyon field in LQC has not been investigated, yet. In this
paper, we discuss the dynamics of tachyon field coupled with a barotropic
perfect fluid in LQC. We focus on the inverse square potential, and try to
explore the possibility of a viable dark energy model.

The organization of this paper is as follows. In Sec.\ref{Sec.2}, we try to derive the
expansion law for a pure tachyonic matter in LQC. In Sec.\ref{Sec.3}, we couple the
field with a barotropic perfect fluid and analyze the dynamics of the
autonomous system. The cosmological implications of the phase plane analysis
are presented in Secs.\ref{Secsec.4A} and \ref{Secsec.4B}. The conclusions are made in Sec. \ref{Sec.5}.

\section{Tachyon Matter in Loop Quantum Cosmology}\label{Sec.2}

\begin{figure}[tbp]
\includegraphics[clip,width=0.45\textwidth]{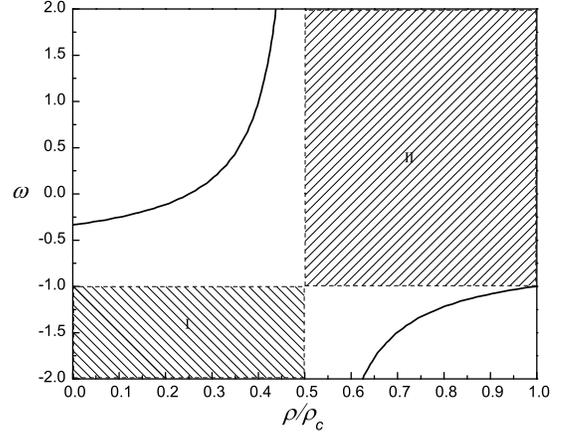}
\caption{$\ddot{a}>0$ in the region between the two solid curves. The region I and II correspond to $\dot{H}>0$.}
\label{Fig-1}
\end{figure}

Based on the holonomy correction in loop quantum cosmology, the modified
Friedmann equation of a flat$(k=0)$ Friedmann-Robertson-Walker (FRW)
cosmological model is
\begin{equation}
H^2=\frac 13\rho \left( 1-\frac \rho {\rho _c}\right) ,  \label{H^2}
\end{equation}
where $H$ is the Hubble parameter, $\rho $ and $\rho _c$ denote the matter
density and critical density, respectively. We also set $8\pi G=1$ for
convenience. The energy conservation equation is the same as the classical
one,
\begin{equation}
\dot{\rho}=-3H\left( \rho +p\right) ,  \label{rho}
\end{equation}
where $p$ is the pressure. Differentiate Friedmann eqaution with respect to
time, we have
\begin{equation}
\dot{H}=-\frac 12\left( p+\rho \right) \left( 1-\frac{2\rho }{\rho _c}%
\right) ,  \label{H}
\end{equation}
where 'dot' denotes the derivative with respect to the cosmological time $t$.
Therefore the conditions for superinflation $(\dot{H}>0)$ are
\begin{equation}
\left\{
\begin{array}{c}
\omega <-1,~~if~~1-\frac{2\rho }{\rho _c}>0, \\
\omega >-1,~~if~~1-\frac{2\rho }{\rho _c}<0.
\end{array}
\right.
\end{equation}
where $\omega =p/\rho $ is the equation of state. It's easy to see the
existence of superinflation in LQC is purely an effect of quantum geometry,
because it originates from the time derivative of the modification term $%
(1-\rho /\rho _c)$. The Raychaudhuri equation then becomes
\begin{equation}
\frac{\ddot{a}}a=\dot{H}+H^2=-\frac 16\left[ 3p\left( 1-\frac{2\rho }{\rho _c%
}\right) +\rho \left( 1-\frac{4\rho }{\rho _c}\right) \right] ,
\end{equation}
which indicates the conditions for $\ddot{a}>0$:
\begin{equation}
\left\{
\begin{array}{c}
w<-\frac 13\frac{1-\frac{4\rho }{\rho _c}}{1-\frac{2\rho }{\rho _c}},~if~1-%
\frac{2\rho }{\rho _c}>0, \\
w>-\frac 13\frac{1-\frac{4\rho }{\rho _c}}{1-\frac{2\rho }{\rho _c}},~if~1-%
\frac{2\rho }{\rho _c}<0.
\end{array}
\right.
\end{equation}
The regions for $\dot{H}>0$ and $\ddot{a}>0$ are portrayed explicitly in
Fig.\ref{Fig-1}. Obviously, as the matter density decreases, the correction term
becomes less and less important, and the Friedmann equation as well as the
conditions for $\ddot{a}>0$ and $\dot{H}>0$ are consistent with classical
cosmology in an asymptotical way.

Now we consider the model with only tachyon field. Note that the tachyon
field referred in this paper is just a scalar field with an nonquadratic
kinetic term. We don't claim any identification with the tachyon in string
theory. According to Sen \cite{Sen-TM,Sen-RT}, the energy density and
pressure of the tachyon field in a flat FRW cosmology can be expressed as
\begin{equation}
\rho _\phi =\frac{V\left( \phi \right) }{\sqrt{1-\dot{\phi}^2}},~p_\phi =-V%
\sqrt{\left( 1-\dot{\phi}^2\right) },
\end{equation}
where $\phi $ is the tachyon field, $V(\phi )$ denotes its potential, and
the equation of state is
\[
\omega _\phi =\frac{p_\phi }{\rho _\phi }=\dot{\phi}^2-1.
\]
We see $\omega _\phi $ ranges smoothly from $-1$ to $0$. Using the energy
conservation Eq. (\ref{rho}), the evolution equation of tachyon field can be
written explicitly as
\[
\ddot{\phi}+\left( 1-\dot{\phi}^2\right) \left( \frac{V^{\prime }}V+3H\dot{%
\phi}\right) =0.
\]
where $V^{\prime }=dV/d\phi $.

Here, we are interested in the inverse square potential
\begin{equation}
V=\beta \phi ^{-2},
\end{equation}
with $\beta >0$. According to \cite{Feinstein} this potential is able to
produce a power law expansion in classical cosmology. In classical
cosmology, one can use the Hubble parameter instead of the field $\phi $ as
a fundamental quantity by employing the Hamilton-Jacobi formulation. By
using Eqs. (\ref{H^2}) and (\ref{H}) and dropping the $\rho /\rho _c$ terms,
one can obtain
\begin{equation}
\dot{\phi}^2=-\frac{2\dot{H}}{3H^2},
\end{equation}
which indicates $\dot{H}\leq 0$, and therefore superinflation will not occur
in classical cosmology. Divide both sides by $\dot{\phi}$, we have
\begin{equation}
\dot{\phi}=-\frac{2H^{\prime }}{3H^2},
\end{equation}
and thus
\begin{equation}
H(\phi )^{\prime }{}^2-\frac 94H(\phi )^4+\frac 14V(\phi )^2=0.
\end{equation}
For the inverse square potential, an exact solution $H\sim \phi ^{-1}$ is
found, and after some algebra one can arrive at
\begin{equation}
\left\{
\begin{array}{c}
a(t)=t^n, \\
\phi (t)=\sqrt{\frac 2{3n}}t,
\end{array}
\right.
\end{equation}
where
\begin{equation}
n=\frac 13+\frac 16\sqrt{4+9\beta ^2}.
\end{equation}
The solution is inflationary when $n>1$, or $\beta >\sqrt{4/3}$. Although
one can have arbitrarily fast expansion with an arbitrarily big $n$ or $%
\beta $, the condition for an accelerated expansion already calls for an
energy scale close to a Plank mass \cite{Cope05,Cope06}. Refs.\cite{sami,
Andrei} also discussed the problem of tachyon model in reheating. Therefore
this tachyon model is more suitable as a dark energy model rather than an
inflaton model.

However, it is difficult to find an exact solution like this in LQC due to
the existence of the quantum modification term. Combining equation (1) and
(3), we have
\begin{equation}
\dot{\phi}^2=-\frac{2\dot{H}}{3H^2}\frac{1-\rho _\phi /\rho _c}{1-2\rho
_\phi /\rho _c}.
\end{equation}
Obviously, $\dot{H}>0$ when $(1-2\rho _\phi /\rho _c)<0$, which means
superinflation will naturally happen and it's purely an effect of quantum
geometry as we said before. Divide both sides by $\dot{\phi}$, we obtain
\begin{equation}
\dot{\phi}=-\frac{2H^{\prime }}{3H^2}\frac{1-\rho _\phi /\rho _c}{1-2\rho
_\phi /\rho _c},
\end{equation}
and therefore
\begin{equation}
H^{\prime }{}^2\left( \frac{1-\rho _\phi /\rho _c}{1-2\rho _\phi /\rho _c}%
\right) ^2-\frac 94H^4+\frac 14V^2\left( 1-\frac{\rho _\phi }{\rho _c}%
\right) ^2=0.
\end{equation}
Fortunately, we are still able to analyze its asymptotical behavior. When $%
\rho _\phi \ll \rho _c$, we can neglect the derivatives of the correction
terms since these terms, since $(1-\rho _\phi /\rho _c)$ and $(1-2\rho _\phi
/\rho _c)$ will not change significantly. Then, an approximated power-law
expansion
\begin{equation}
a(t)\sim t^m
\end{equation}
can be obtained, with
\begin{equation}
m=\frac 16\left( \frac{1-\rho _\phi /\rho _c}{1-2\rho _\phi /\rho _c}\right) \left[
2+\sqrt{4+\frac{9\beta ^2\left( 1-2\rho _\phi /\rho _c\right)
^4}{\left( 1-\rho _\phi /\rho _c\right) ^2}}\right] .
\end{equation}
So we have
\begin{eqnarray}
\phi (t)&=&\sqrt{\frac{1-\rho _\phi /\rho _c}{1-2\rho _\phi /\rho _c}\frac 2{3m}}t  \nonumber \\
&=&\frac{2t}{\sqrt{2+\sqrt{4+\frac{9\beta ^2\left( 1-2\rho _\phi /\rho
_c\right) ^4}{\left( 1-\rho _\phi /\rho _c\right) ^2}}}}.
\end{eqnarray}
Obviously, $m\rightarrow n$ as $\rho _\phi /\rho _c\rightarrow 0$. The
quantum geometry results in a different evolution of tachyon field and the
scale factor,but the evolution will converge to the classical one at late
time when the quantum effects vanishes.

\section{With Barotropic Fluid}\label{Sec.3}

In order to obtain a viable dark energy model, we add a barotropic perfect
fluid in our model, for which the equation of state is $p_\gamma =(\gamma
-1)\rho _\gamma $, where $\gamma $ is a constant.Therefore, the equation of
state for the whole is
\begin{eqnarray}
\omega  &=&\frac p\rho =\frac{\omega _\phi \rho _\phi +\omega _\gamma \rho
_\gamma }\rho   \nonumber \\
&=&\omega _\phi \Omega _\phi +\omega _\gamma \Omega \gamma =(\phi ^2-\gamma
)\Omega _\phi +\gamma -1.
\end{eqnarray}
where
\[
\omega _\gamma =\frac{p_\gamma }{\rho _\gamma }=\gamma -1,
\]
and $\Omega _\phi $ and $\Omega _\gamma $ are the fractional densities
defined as
\begin{equation}
\Omega _\phi =\frac{\rho _\phi }\rho ,\ \Omega _\gamma =\frac{\rho _\gamma }%
\rho =1-\Omega _\phi .
\end{equation}
If we simply assume there is no interaction between the tachyon field and
barotropic fluid, then their evolution equations are
\begin{equation}
\ddot{\phi}+\left( 1-\dot{\phi}^2\right) \left( \frac{V^{\prime }}V+3H\dot{%
\phi}\right) =0,  \label{evolution-eqs-1}
\end{equation}
and
\begin{equation}
\dot{\rho}_\gamma +3\gamma H\rho _\gamma =0.  \label{evolution-eqs-2}
\end{equation}
Eq.(\ref{H}) now becomes
\begin{equation}
\dot{H}=-\frac 12\left( \dot{\phi}^2\rho _\phi +\gamma \rho _\gamma \right)
\left( 1-2\frac \rho {\rho _c}\right) .  \label{H2}
\end{equation}
Combining Eqs. (\ref{evolution-eqs-1})-(\ref{H2}) and the Friedmann
equation, we can construct a 4-dementional autonomous system. To see this,
we usually introduce 4 convenient variables \cite{juan,kui}:
\begin{equation}
\left\{
\begin{array}{c}
x\equiv \dot{\phi},\ y\equiv \sqrt{\frac V{3H^2}}, \\
z\equiv \frac \rho {\rho _c},\ \lambda \equiv \frac{V^{\prime }}{3HV}.
\end{array}
\right.   \label{xyzl-1}
\end{equation}
Moreover, we use $N=\ln a^3$ instead of the cosmological time $t$ as an
independent variable, therefore for any time-dependent function $f$, we have
\begin{equation}
\frac{df}{dN}=\frac{\dot{f}}{3H}.
\end{equation}
Here, one should be careful that this treatment may not be practicable when $%
\dot{a}=0$, however, at that point, one can always switch back to $t$
without causing any problem. With the help of new variables, the Eqs. (\ref
{H^2}) and (\ref{evolution-eqs-1})-(\ref{H2}) can now be expressed
respectively as follows:
\begin{equation}
\left\{
\begin{array}{c}
\left( \frac{\rho _\gamma }{3H^2}+\frac{y^2}{\sqrt{1-x^2}}\right) \left(
1-z\right) =1, \\
\ddot{\phi}=-\left( 1-x^2\right) \left( \frac{V^{\prime }}V+3Hx\right) , \\
\frac{\dot{\rho _\gamma }}{3H^3}=-3\gamma \left( \frac 1{1-z}-\frac{y^2}{%
\sqrt{1-x^2}}\right) , \\
\frac{\dot{H}}{3H^2}=-\frac {\left(1-2z\right)}2\left[ \frac{x^2y^2}{\sqrt{1-x^2}}+\gamma
\left( \frac 1{1-z}-\frac{y^2}{\sqrt{1-x^2}}\right) \right] .
\end{array}
\right.
\end{equation}

Using the above equations, differentiate $x$, $y$, $z$, and $\lambda $ with
respect to $N$, the nonlinear dynamical system is shown in an obvious
autonomous form:
\begin{equation}
\left\{
\begin{array}{c}
\frac{dx}{dN}=-\left( 1-x^2\right) \left( x+\lambda \right) , \\
\frac{dy}{dN}=\frac {y\left(1-2z\right)}2\left\{ x\lambda +\left[ \frac{x^2y^2}{\sqrt{1-x^2}}%
+\gamma \left( \frac 1{1-z}-\frac{y^2}{\sqrt{1-x^2}}\right) \right] \right\} , \\
\frac{dz}{dN}=-z\left[ x^2+\left( \frac 1{1-z}-\frac{y^2}{\sqrt{1-x^2}}%
\right) \left( \gamma -x^2\right) \left( 1-z\right) \right] , \\
\frac{d\lambda }{dN}=x\lambda ^2\left( \Gamma -1\right) +\frac {\lambda \left( 1-2z\right)} 2%
\left[ \frac{x^2y^2}{\sqrt{1-x^2}}+\gamma \left( \frac 1{1-z}-\frac{y^2}{%
\sqrt{1-x^2}}\right) \right] ,
\end{array}
\right.   \label{xyzl-2}
\end{equation}
where
\begin{equation}
\Gamma =\frac{V^{\prime \prime }}{V^{\prime }{}^2},  \label{Gamma}
\end{equation}
can be fixed if the potential is specified. Note that, the autonomous system
in LQC has one more dimension than the one in classical cosmology, because
of the modification term $\rho /\rho _c$ \cite{kui}. For the inverse square
potential $V(\phi )=\beta \phi ^{-2}$ investigated in this paper, $\Gamma
=3/2$. Moreover, we find $\lambda =-\alpha y$ for this potential, where
\begin{equation}
\alpha \equiv \sqrt{\frac 4{3\beta }},
\end{equation}
which means the system can be reduced into a 3-dimensional one:
\begin{equation}
\left\{
\begin{array}{c}
\frac{dx}{dN}=-\left( 1-x^2\right) \left( x-\alpha y\right) , \\
\frac{dy}{dN}= -\frac{\alpha xy^{2}}2+\frac {\left(
1-2z\right)y}2\left[ \frac{x^2y^2}{\sqrt{1-x^2}}%
+\gamma \left( \frac 1{1-z}-\frac{y^2}{\sqrt{1-x^2}}\right) \right], \\
\frac{dz}{dN}=-z\left[ x^2+\left( \frac 1{1-z}-\frac{y^2}{\sqrt{1-x^2}}%
\right) \left( \gamma -x^2\right) \left( 1-z\right) \right] .
\end{array}
\right.   \label{xyz}
\end{equation}

Before we analyze the autonomous system described by Eq.(\ref{xyz}), we
should state several physical restrictions. First, the density should be
limited and real-valued, therefore $-1<x<1$. The potential is required to be
positive, so we have $y>0$. In the analysis, we add the three end points of $%
x$ and $y$ for convenience, i.e. we set $x\in [-1,1]$, $y\in [0,+\infty ]$.
Moreover, the total density should be nonnegative and lower than the
critical density, i.e. $0\leq z\leq 1$. Moreover, we apply a normal
restriction for $\gamma $, that is $0<\gamma <2$. Therefore, $\omega $ will
range from $-1$ to $1$ due to the choice of $\gamma $. Since neither of the
two parts should have a negative density, we have $\Omega _\phi ,\Omega
_\gamma \in [0,1]$. This restriction leads to $(1-z)^2y^4+x^2\leq 1$, and
the equal sign is for $\Omega _\phi =1$. This indicates the physically
possible phase plane projection on the $x-y$ plane shrinks as $z$ decreases
from 1 to 0. In fact, as $z\rightarrow 0$ the phase plane projection becomes
$y^4+x^2\leq 1$ asymptotically.

The equilibrium points or fixed points $(x_e,y_e,z_e,\lambda _e)$ are
solutions acquired by setting
\begin{equation}
\frac{dx}{dN}=\frac{dy}{dN}=\frac{dz}{dN}=0.
\end{equation}
According to Lyapunov's theory of stability, the stability of a fixed point
can be determined by the property of the linearized system about it. The
linearization is done by expanding  Eq.(\ref{xyz}) about the fixed points
and keeping only the linear parts. After that, one can obtain a matrix
\begin{equation}
\left(
\begin{array}{lll}
\frac \partial {\partial x}\frac{dx}{dN} & \frac \partial {\partial y}\frac{%
dx}{dN} & \frac \partial {\partial z}\frac{dx}{dN} \\
\frac \partial {\partial x}\frac{dy}{dN} & \frac \partial {\partial y}\frac{%
dy}{dN} & \frac \partial {\partial z}\frac{dy}{dN} \\
\frac \partial {\partial x}\frac{dz}{dN} & \frac \partial {\partial y}\frac{%
dz}{dN} & \frac \partial {\partial z}\frac{dz}{dN}
\end{array}
\right) _{(x=x_e,y=y_e,z=z_e)}  \label{matrix}
\end{equation}
at each fixed point. If the all the eigenvalues of (\ref{matrix}) for a
fixed point have negative real parts, the fixed point is (locally)
exponentially stable. However, if there exist one or more eigenvalues which
have positive real parts, then the fixed point is unstable. In both stable
and unstable cases, the fixed point is a node if the eigenvalues are all
real, otherwise it will be a spiral point. Furthermore, the fixed point is
called saddle point if the eigenvalues have both positive and negative real
parts.

We found up to 5 fixed points for Eq.(\ref{xyz}) , their properties are
listed in Table I, where
\begin{equation}
y_1=\sqrt{\frac{\sqrt{\alpha ^4+4}-\alpha ^2}2},0<\alpha y_1<1.
\end{equation}

The physical requirements mentioned before also set restrictions on the
existence and stability of these fixed points.By considering $\gamma /\alpha
^2/\sqrt{1-\gamma }\leq 1$, we found
\begin{equation}
0\leq \frac{\gamma (\gamma -2)^2}{\sqrt{1-\gamma }}\leq f(\gamma)\leq \alpha
^2(2-\gamma )^2,
\end{equation}
where $f(\gamma)=4\alpha^2-20\gamma\alpha^2+17\gamma^2\alpha^2+16\gamma^2\sqrt{1-\gamma}$,
which indicates the two eigenvalues of $P_3$ with seemingly complicated
square root part are definitely real and nonpositive.
Note that, when $\gamma >\alpha ^2y_1^2$, $P_3$ doesn't exist. There are only four fixed
points in the system, and $P_4$, as a stable node, is the only attractor in
the system. When $\gamma =\alpha ^2y_1^2$, a bifurcation occurs as the fifth
fixed point, $P_3$, emerges and coincides with $P_4$. As $\gamma $ decreases
from $\alpha ^2y_1^2$ to $0$, $P_4$ turns into a unstable saddle and $P_3$
becomes a new attractor. The position of $P_3$ is located on a straight line
on the x-y plane described by $(x,y)=(\sqrt{\gamma },\sqrt{\gamma }/\alpha )$
and in the limit case $\gamma =0$, it will coincide with $P_1$.

\section{Late Time Evolution}\label{Sec.4}

\begin{figure}[tbp]
\includegraphics[clip,width=0.45\textwidth]{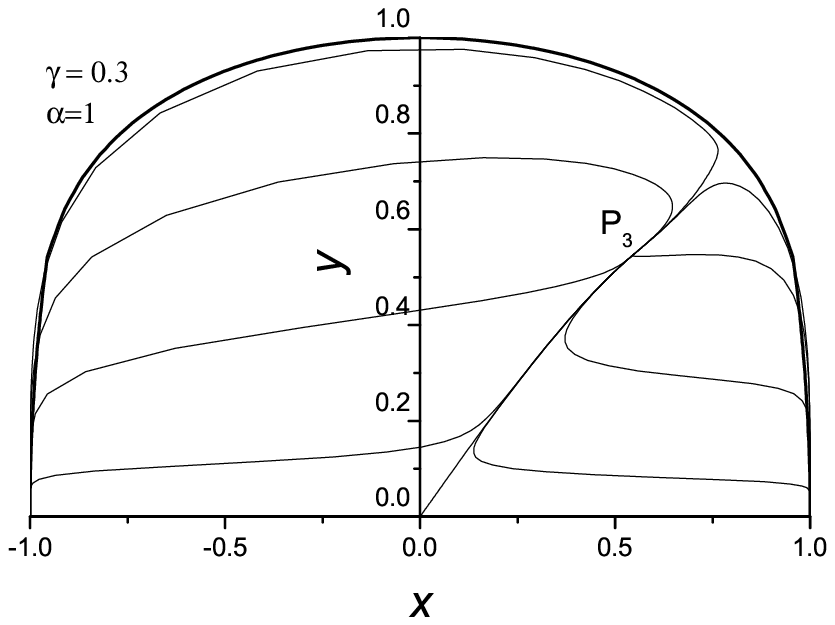}
\caption{Phase plane for $\gamma =0.3,\alpha =1$. The outer contour
corresponds to $y^4+x^2=1$. Solutions start at z=0.01. Nearly all solutions
end up at $P_3$.} \label{Fig-2}
\end{figure}

\begin{figure}[tbp]
\includegraphics[clip,width=0.45\textwidth]{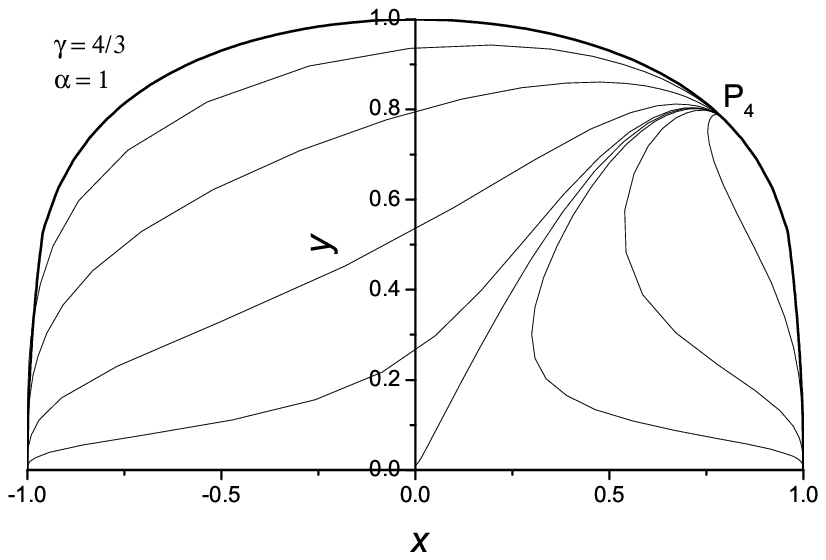}
\caption{Phase plane for $\gamma =4/3,\alpha =1$. The outer contour
corresponds to $y^4+x^2=1$. Solutions start at z=0.01. Almost all solutions
end up at $P_4$.} \label{Fig-3}
\end{figure}

From TABLE I we can see that all the fixed points locate at $z=0$, where the
energy density vanishes and the LQC modification term $\rho /\rho _c$
becomes unimportant. We have proved in last section that the the physically
possible phase plane projection on $x-y$ plane becomes $y^4+x^2\leq 1$
asymptotically. Therefore for the late time evolution, it is sufficient for
us to investigate the phase projection on the $x-y$ plane at small $z$ by
noting that $z$ monotonously decreases as $a$ or $N=\ln a^3$ increases,
which we can easily read from Eq. (\ref{Gamma}). In the previous section, we
have proved that there is at most one stable point in our model, then nearly
all solutions will end there (see Figs.\ref{Fig-2} and \ref{Fig-3}). This attracting behavior allows
us to investigate the properties of late time evolution near the attractor,
regardless of the history before it. The fine-tuning problem can thus be
waived since the same ending occurs for a wide range of initial conditions.
Here we analyze two different solutions, the tachyon dominated solutions and
tracker solutions.
\begin{widetext}
\begin{center}
\begin{table}[tbp]
\caption{Properties of fixed points}
\label{Table I}
\begin{tabular}[c]{ccccccc}
\hline\hline
Points & Coordinates & Existence & Eigenvalue & $\Omega_\phi$ & Stability \\ \hline
$P_1$ & $\left(0,0,0\right)$ & for all $\gamma$ & $\left(-1,-\gamma,\frac{1}{2}%
\gamma\right)$ & 0 & Unstable saddle \\ \hline
$P_{2\pm}$ & $\left(\pm1,0,0\right)$ & for all $\gamma$ & $\left(2,-\gamma,\frac{1%
}{2}\gamma\right)$ & 1 & Uunstable saddle \\ \hline
$P_3$ & $\left(\sqrt{\gamma},\frac{\sqrt{\gamma}}{\alpha},0\right)$ & $0 <
\gamma\leq\alpha^2y_1^2$ & $\left(-\gamma,\frac{-\left(2-\gamma\right)\alpha%
-\sqrt{f(\gamma)}}{4\alpha},\frac{-\left(2-\gamma\right)\alpha%
+\sqrt{f(\gamma)}}{4\alpha}\right)$ & $\frac{\gamma/\alpha^2}{\sqrt{1-\gamma}}$ &
Stable node, if $0 <\gamma<\alpha^2y_1^2$ \\
&&&where $f(\gamma)=4\alpha^2-20\gamma\alpha^2+17\gamma^2\alpha^2+16\gamma^2\sqrt{1-\gamma}$&&\\ \hline
$P_4$ & $\left(\alpha y_1,y_1,0\right)$ & for all $\gamma$ & $\left(-\alpha^2{y_1}%
^2,-\frac{1}{2}\left(2-\alpha^2{y_1}^2\right),-\left(\gamma-\alpha^2{y_1}%
^2\right)\right)$ & 1 & Stable node, if $\gamma>\alpha^2{y_1}^2$;\\
&&&&&Unstable saddle, if $\gamma<\alpha^2{y_1}^2$ \\
\hline\hline
\end{tabular}
\end{table}
\end{center}
\end{widetext}


\subsection{Tracker solutions}\label{Secsec.4A}

When $\gamma <\alpha ^2y_1^2$, $P_4$ becomes an unstable saddle while $P_3(%
\sqrt{\gamma },\sqrt{\gamma }/\alpha ,0)$ comes out as the only attractor in
our system. The fractional density is a constant at $P_3$, which means both
matter scale as the same power of $a$. Therefore, we have
\begin{equation}
\Omega _\phi \approx \frac{\gamma /\alpha ^2}{\sqrt{1-\gamma }},\ \omega
_\phi \approx \omega _\gamma =\gamma -1,
\end{equation}
\begin{equation}
\phi (t)\approx \sqrt{\gamma }t+\phi _0,
\end{equation}
\begin{equation}
\rho _\gamma \propto \rho _\phi \propto a^{-3\gamma },~\ a(t)\propto t^{%
\frac 2{3\gamma }},
\end{equation}
near the attractor, where $\phi _0$ is an integration constant. For a same $%
\beta $ or $\alpha $, this solution produces a faster expansion than the one
with tachyon as the only matter. It will lead to an eternal acceleration if $%
\gamma <\min \{2/3,\alpha ^2y_1^2\}$, while if $2/3<\gamma <\alpha ^2y_1^2$,
a deceleration phase will definitely occur at late time. Due to the choice
of $\gamma $ and $\alpha $, $\Omega _\phi $ can range from 0 to 1. However,
Refs. \cite{Cope05,Cope06} argues that this solution cannot be a viable one,
since its existence requires $\gamma <\alpha ^2y_1^2<1$.

\subsection{Tachyon dominated solutions}\label{Secsec.4B}

In the case $\gamma >\alpha ^2y_1^2$, $P_4$ does not exist. As the solutions
converge to the only attractor $P_3(\alpha y_1,y_1,0)$ in late time
evolution, tachyon will become dominant while both matters are decreasing.
In fact, near the attractor we will have
\[
\Omega _\phi \approx 1,~~\rho _\phi \approx \rho \ll \rho _c,~~\omega
\approx \omega _\phi \approx \alpha ^2y_1^2-1,
\]
\begin{equation}
\phi (t)\approx \alpha y_1t,
\end{equation}
\[
\rho _\phi \propto a^{-3\left( 1+\omega _\phi \right) }=a^{-3\alpha
^2y_1^2},~a(t)\propto t^{2/3\left( 1+\omega _\phi \right) }=t^{2/3\alpha
^2y_1^2}.
\]
This result is consistent with the result we derived in the second section.
It is easy to see that if $\alpha ^2y_1^2<2/3$, the solutions depict an
eternal acceleration. On the other hand, the universe will end up in
deceleration if $\alpha ^2y_1^2>2/3$.

An interesting possibility is to see tachyon field as a combination of two
parts \cite{Pad02,Pad03}, which behave like a pressureless dust (dark
matter, denoted by lower case $DM$) and a cosmological constant(dark energy,
denoted by lower case $\Lambda $), respectively:
\begin{equation}
\rho _\phi =\rho _{DM}+\rho _\Lambda ,\ p_\phi =p_{DM}+p_\Lambda ,
\end{equation}
where
\begin{equation}
\rho _{DM}=\frac{V(\phi )\dot{\phi}^2}{\sqrt{1-\dot{\phi}^2}},~p_{DM}=0,
\end{equation}
and
\begin{equation}
\rho _\Lambda =V(\phi )\sqrt{1-\dot{\phi}^2},~p_\Lambda =-\rho _\Lambda .
\end{equation}
In this way, dark matter and dark energy originate from a same scalar field,
and the dynamics of tachyon field becomes a description of their dynamics
and interaction. The ratio between the two parts is
\begin{equation}
\frac{\rho _{DM}}{\rho _\Lambda }=\frac{\dot{\phi}^2}{1-\dot{\phi}^2}.
\end{equation}
The proportion of dark matter rises as $\dot{\phi}^2$ increases. If the
barotropic fluid is radiation$(\gamma =4/3)$, then, by the virtue of the
attractor solution, it is possible to have a trajectory that goes from the
radiation dominated era to the matter dominated era, and then to the dark
energy dominated regime. Radiation should dominate first, that is $\Omega
_\gamma \approx 1$. Then, to have the universe dominated by matters
described by tachyon field after the era dominated by radiation, that is $%
\Omega _\phi \approx 1$, we need the trajectory to stay close to the
boundary of the phase plane, i.e. $(1-z)^2y^4+x^2\approx 1$, or $%
y^4+x^2\approx 1$ because $z$ is very small. To have a sufficient long
matter dominated era before dark energy take over, we just need the
trajectory to stay close to the saddle point $P_{2\pm }$ after radiation's
domination. Therefore we will rule out the trajectories which are far away
from the boundary of the phase plane in following discussion.

The negative/positive branch, which starts near $P_{2-}$/$P_{2+}$, will show
a quite different evolution, because the attractor only lies in the right
half of the phase plane and therefore the negative branch will have to get
across to arrive at the attractor. The cosmological consequence will also be
quite different. For the positive branch, the ratio of dark energy will
increase as $\dot{\phi}$ decreases. If the attractor is inflationary ($%
\alpha ^2y_1^2<2/3$), the allowed trajectories can go naturally into the
dark energy dominated regime that leads to acceleration. While if the
attractor is not inflationary($\alpha ^2y_1^2>2/3$), acceleration will not
take place because $\dot{\phi}$ decreases almost monotonously for the
feasible trajectories which are close to the boundary. On the other hand,
for the negative branch, the universe will definitely enter a dark energy
dominated regime that leads to acceleration when $\dot{\phi}^2<2/3$, or $%
-2/3<\dot{\phi}<2/3$, since the trajectories will get across from the left
half to the right half. Note that, for the negative branch, the universe is
not expanding in the power-law way we described before when it first gets
into the acceleration regime $-2/3<\dot{\phi}<0$. When $\dot{\phi}=0$, dark
energy dominates completely. After that, the ratio of matter increases
again, and the final state of universe will(or will not) be inflationary if $%
\alpha ^2y_1^2<2/3$(or $\alpha ^2y_1^2>2/3$). So there is possibility that
we are just currently living in a transitory accelerating period and the
acceleration rate can change according to the dynamics of the tachyon field.

\section{Conclusion}\label{Sec.5}

Previous works on tachyon cosmology in spatially flat FRW universe showed a
purely tachyonic matter with an inverse square potential $V=\beta \phi^{-2}$
leads to a power-law expansion \cite{Feinstein}. In LQC scenario, although
it is hard to find a exact solution, the expansion of universe is nearly a
power-law one when $\rho/\rho_c$ is small. When the tachyon field is coupled
with a barotropic perfect fluid with $0<\gamma<2$, we are able to find two
kinds of stable nodes which exist exclusively to each other and represent
different cosmological situations. The tracker solution exists for $%
\gamma<\alpha^2 y_1 ^2$, while the tachyon dominated solution exists for $%
\gamma>\alpha^2 y_1 ^2$. Refs.\cite{Cope05,Cope06} argued that the tracker
solution cannot be a viable one, since its existence requires $%
\gamma<\alpha^2y_1^2<1$. Therefore we focused on the tachyon dominated
solution. We considered the tachyon field as a combination of two parts
which respectively behave like dark matter and dark energy. Because of the
existence of stable node(attractor), we found when $\gamma=4/3$, it's
possible to have a trajectory in which the universe evolves naturally from
radiation dominated regime to matter dominated regime, and then into the
current dark energy dominated regime through qualitative discussion. The
negative and positive branches, which respectively start from close to $\dot{%
\phi}<0$ and $\dot{\phi}>0$, can be interpreted into different cosmological
evolutions and it is possible that we are just living in a transitory
accelerating period, while the final stage of universe will be identical for
the same $\alpha$ or $\beta$. However, this perspective to see tachyon field
as a combination of two parts revives the need of fine tuning. We also
avoided the discussion of the period before radiation dominated era. Further
work can be done to bridge the gap.

\acknowledgments This work was supported by the National Natural Science Foundation of China (Grant Nos. 11175019 and 11235003) and the Fundamental Research Funds for the Central Universities.

\end{document}